# Laser-induced fluorescence based characterisation method for aggregation behaviour of rhodamine B (RhB) in water, ethanol, and propanol

Seikh Mustafa Radiul[a], Jugal Chowdhury[b], Angana Goswami[c], Simanta Hazarika[a,*]

[a] Department of Physics, Gauhati University, Assam, India

[b] Institute for Plasma Research, Gandhinagar, Gujarat, India

[c] Department of Physics, Pandit Deendayal Upadhyaya Adarsha Mahavidyalaya, Dalgaon, Assam, India

*Corresponding Author Email Id: simantahazarikagu@gmail.com

#ORCID iD: https://orcid.org/0000-0003-1012-3066

**Abstract:**

The aggregation behaviour of rhodamine B (RhB) dye has been studied with the change in the concentration of the RhB in water, ethanol and propanol using absorption and laser induced fluorescence spectroscopy. The dimer and monomer fluorescence emissions were observed simultaneously in all the solvents. The monomer to dimer fluorescence intensity ratio $\left(\frac{I_M}{I_D}\right)$, where $I_M$ = Monomer fluorescence intensity, $I_D$ = Dimer fluorescence intensity, has been calculated and found that at a certain concentration of RhB the magnitude of the ratio is different in different solvents. The ratio $\left(\frac{I_M}{I_D}\right)$ becomes equivalent to one for 1.5 gm/l, 3 gm/l and 4.2 gm/l, (gm/l: gram/litre) concentration of RhB in water, ethanol and propanol respectively. This concentration of RhB at which the ratio $\left(\frac{I_M}{I_D}\right)$ become unity is proposed as "*critical concentration*". Above this critical concentration the dimeric fluorescence dominates and below which the monomeric fluorescence dominates. The parameter "*critical concentration*" provides a condition of equilibrium between the monomer and their aggregates. Moreover, in the solvent for which the magnitude of *critical concentration* is less the fluorophore molecule aggregates more easily and vice versa. The ratio $\left(\frac{I_M}{I_D}\right)$ and *critical concentration* quantify the aggregation process of the fluorophore in the solvent. Hence, the "*critical concentration*" value of the fluorophore (RhB) could be assumed as the characteristic parameter to study the aggregation behaviour of the fluorophore in solvents. The described method has the merit of use in the study of lasing ability and the wavelength tunability of the dye laser gain media as well as in the heavy metal sensing technology in water. This method may also be extended for other fluorophores and solvents.

**Keywords:** Lase Induced Fluorescence, Aggregation, Dimer, Monomer, Rhodamine B, Critical-concentration,

**Introduction:**

Xanthene dyes are most effectively and extensively explored dye for their wide application in various fields from fluorescence microscopy in biotechnology to cost-effective in-situ sensing technology [1]. Rhodamine B (RhB) is a special class of fluorescent xanthene dye that emits in the red wavelength region [2]. It is mostly favoured for its excellent photostability, photophysical properties, and high fluorescence yield property that make it suitable as laser dye and fluorescent sensing probes for the detection process [3].

Most of the rhodamine class of xanthene dye such as rhodamine 6G, rhodamine B, etc. undergo self-aggregation depending on its concentration, structure and the types of solvent used [4]. The aggregation of rhodamine-B dye in solutions leads to the formation of dimer species [5]. These dimers show fluorescence as well as phosphorescence [5, 6]. Excimer is also formed by the association of two molecules one of which is excited. On the contrary, dimerization is a process of aggregation of molecules of the same kind in their ground state [7]. The absorption spectrum of the excimer is the same as the monomer absorption spectrum as the excimer is formed after the absorption of the excitation energy [8]. The absorption spectrum of the solution of the molecules will be different with and without the aggregation which signifies the formation of dimer [9]. Thus the emission spectrum also changes, as the fluorescence of the dimer generally occurs from the lower energy levels of the excited states than that of the excited state fluorescence of the monomer [10].

The emission spectrum of a dye normally covers a wavelength range between 30 nm and 50 nm. Therefore, a careful selection from the available laser dyes permits the design of tunable lasers from 320 nm to 1200 nm [11]. Organic dyes have been widely used as a gain medium for optofluidic ring resonators (OFRR) lasers. Rhodamine B dye is a potential candidate for the laser gain media because of its high fluorescence yield and its conjugated structure i.e. alternate single and double bonds [12]. Rhodamine B is usually selected as a gain medium for OFRR laser in the wavelength range of 578-610 nm and it is generally dissolved in solvents such as pure water, ethanol, ethylene glycol, and glycerol [13]. The lasing threshold, the tuning range of the emitted lasing wavelength, and the fluorescence quantum yield of the dye directly depend on the solvent of the dye [14, 15]. Therefore, the selection of the solvent for the dyes in the laser gain media is greatly essential. It has been reported earlier for rhodamine 6G that at higher aggregates generally the quantum yield of the dimeric form of the dye is low [16], and that degrades the lasing ability of the dye system.

The dimer state of the rhodamine-B has been reported earlier [17, 18]. Most of the papers explain the formation of dimer species by analysing the change in the absorption spectrum of the rhodamine-B dye. Arbeloa et al. [17] reported the dimeric state of cationic rhodamine-B dye by calculating the aggregation constant from the absorption spectra of a concentrated solution in ethanol. Chambers et al. [19] described Polarized emission and excitation spectra of rhodamine dyes to examine the relative polarizations of both the monomer and dimer bands in its absorption spectra. Gal et al. [20] explain a model for analyzing the absorption spectrum of dimeric states in terms of exciton coupling of the vibronic levels of the monomeric rhodamine B molecules. Arbeloa et al. [21] reported the aggregation of neutral and cationic rhodamine B molecules and analyzed the absorption spectra of its dimers. Mchedlov-Petrosyan et al. [18] also reported the aggregation of rhodamine B in water. Only a few reports explain the effect of the formation of dimer species of rhodamine-B solution over fluorescence [22-24]. Furthermore, no report is available in the literature where fluorescence emission of the monomeric and dimeric forms of rhodamine B is observed simultaneously.

Herein we report an experimental study of the formation of the dimeric state of rhodamine-B simultaneously with its monomeric state in three different solvents (water, ethanol, and propanol). The monomer and dimer fluorescence spectra were studied with the change in the concentration of the rhodamine-B dye in the solutions. This study explains a novel quantitative method based on the equilibrium condition of monomeric and dimeric form of rhodamine B to study its aggregation process in different solvents.

**Materials and method:**

*Materials:*

The Xanthene dye rhodamine-B used was purchased from Sigma Aldrich Co. USA and was used as received. The solvent used in the experiment is deionized water collected from the departmental laboratory and ethanol and propanol were purchased from Sigma Aldrich Co. USA.

*Preparation of the solution:*

Solutions of rhodamine B dye of varying concentrations were prepared separately in water, ethanol, and propanol. To study the effect of concentration over fluorescence:

1) four different solutions of rhodamine B of varying concentrations (0.5 gm/l, 1.0 gm/l, 1.5 gm/l, 2.0 gm/l) in deionized water were prepared,

2) eight different solutions of rhodamine B with concentrations 0.5 gm/l, 1 gm/l, 2 gm/l, 3 gm/l, 5 gm/l, 8 gm/l, 10 gm/l and 12 gm/l were prepared in ethanol and,

3) thirteen different solutions of rhodamine B dye of concentrations 0.5 gm/l, 1.5 gm/l, 2.5 gm/l, 3.5 gm/l, 4.5 gm/l, 5.5 gm/l, 6.5 gm/l, 8.0 gm/l, 10 gm/l, 12.0 gm/l, 15gm/l, 18 gm/l and 25 gm/l were prepared in propanol. The gm/l stands for gram/litre.

*Experimental setup:*

A 405 nm collimated diode laser system (Laserglow Technologies) of average power 118.5 mW was used as an excitation source for the LIF (Laser-induced Fluorescence) experiment. A concave grating spectrometer of wavelength range 320 nm to 900 nm and resolution of 0.9 nm, equipped with Toshiba linear CCD array detector and data acquisition and display software (spectra analyte V2.26), were used to record the spectra. The 405 nm laser beam was incident to excite the solution in a quartz cuvette and the emission spectra were collected approximately at $90^0$ to the incident beam with the help of an adjustable slit attached to the spectrometer. For the assurance of minimal contribution of reflected light in the spectra, an angle of $30^0$ is maintained between the incident laser beam and the plane of the cuvette holding the sample. The whole experimental setup was kept fixed with the utmost caution to record the spectra of all the samples.

The UV visible absorption spectra of rhodamine B in water, ethanol, and propanol were recorded in Shimadzu UV-VIS spectrophotometer (model: UV-1900i) with scanning range from 200 nm to 1100 nm, available at departmental instrumentation facility. A very low concentrated solution of rhodamine B in water was initially considered to record the absorption spectrum. The concentration of rhodamine B solution was then progressively increased until a change in the absorption spectrum was observed. The same method is used to record the absorption spectrum of rhodamine B in ethanol and propanol. All the experiments were carried out at a temperature in the range $28^0$-$30^0$ C.

**Results and Discussions:**

*UV visible absorption spectra:*

The absorption spectra of rhodamine B, each in water, ethanol and propanol are shown in figure 1 for both dilute and concentrated solutions. The absorption spectra of a dilute solution of rhodamine B in all three solvents show an intense peak at wavelength 553 nm with a negligible hump in the lower wavelength side. This peak is generally assigned to the monomeric form of rhodamine B [5]. The absorption spectrum of the concentrated solution of rhodamine B is found to be different from that of its dilute solution. In the concentrated solution of all the solvents along with the regular absorption maxima at 553 nm, a second absorption peak with higher intensity is observed at a shorter wavelength (515 nm). This absorption at a shorter wavelength (515 nm) results in the intense peak attributed to the dimer absorption [25, 26].

*Fluorescence spectra of rhodamine B (RhB):*

*In water:*

The fluorescence emission spectra of RhB solution in water for different concentrations (0.5 gm/l, 1.0 gm/l, 1.5 gm/l, and 2.0 gm/l) are shown in figure 2. The fluorescence spectra reveal two distinct peaks for all the solutions of RhB with different concentrations except the solution with the lowest concentration of 0.5 gm/l. The wavelength of the first intense peak varies from 608 nm to 630 nm as the concentration of the RhB solution increases in water. This intense peak at the lower wavelength side is assigned to the monomeric form of RhB [23]. It can be seen that the monomer peak is red-shifted as the concentration of the RhB solution increases. It is to be noted that a similar red-shift effect was earlier reported for methanol [24]. The fluorescence peak observed at the higher wavelength side with increasing concentration is due to the aggregation of the molecules to form dimers at higher concentrations and it appears at 685 nm for all the RhB solutions in water. At the lowest concentration (0.5 gm/l) of RhB in water the dimer fluorescence peak is missing. With the increase of concentration of the RhB in water the intensity of the monomer peak is decreasing and the dimer peak is increasing. When the concentration of RhB was increased further (more than 2.0 gm/l) the incident laser radiation was completely absorbed and very weak fluorescence emission was observed. At this highest concentration, the fluorescence radiation disappeared due to self-quenching or reabsorption [27].

The ratio of the monomer to dimer fluorescence emission peak intensity $\left(\frac{I_M}{I_D}\right)$ of the rhodamine B solution in water with respect to different concentrations of RhB in the solution is shown in table 1. The plot of the monomer to dimer ratio against the concentration of RhB in water is shown in figure 3. This plot follows a perfect power equation,

$$y = 1.5859 \, x^{-1.14} \quad \text{with } R^2 = 0.9988 \tag{1}$$

where x and y represent the concentration of RhB in water and monomer to dimer fluorescence emission peak intensity ratio respectively. $R^2$ is a statistical measure which quantifies the goodness of fit of the model.

*In ethanol:*

The fluorescence emission spectra of RhB solution in ethanol with concentration 0.5 gm/l, 3 gm/l, 5 gm/l, 8 gm/l and 12 gm/l is depicted in figure 4. Two intense peaks are observed for each RhB solution except for the solution with the lowest concentration of 0.5 gm/l. The first intense peak with wavelength ranging from 590 nm to 620 nm with increasing concentration of the RhB in the solution is the signature of emission of its monomeric form in the spectra. The second intense peak observed at around 670 nm is due to the emission of the dimeric form of RhB. As the concentration of RhB in ethanol increases the intensity of the monomer peak gradually decreases and is about to extinct at the highest concentration 12 gm/l. The fluorescence spectra of RhB solution of concentration 1 gm/l, 2 gm/l, and 10 gm/l were found to be similar to that of 3 gm/l and 8 gm/l respectively, and hence not shown in figure 4.

The monomer and dimer fluorescence peak intensity is observed from the fluorescence emission spectrum corresponding to different concentrations of RhB in the solution and is shown in table 2. The monomer to dimer fluorescence emission peak intensity ratio corresponding to different concentrations of RhB is also calculated and shown in table 2. The monomer to dimer fluorescence emission peak intensity ratio is plotted against the different concentrations of RhB in the solution and shown in figure 5. It is found that this plot follows a power equation,

$y = 2.1991\ x^{-0.741}$ with $R^2 = 0.9417$ (2)

The corresponding x and y in the above equation represent the concentration of RhB in the solution and monomer to dimer fluorescence emission peak intensity ratio respectively. $R^2$ is a statistical measure which quantifies the goodness of fit of the model.

### *In propanol:*

The fluorescence spectra of RhB solution in propanol for different concentrations (0.5 gm/l, 1.5 gm/l, 2.5 gm/l, 5.5 gm/l, 8.0 gm/l, 12.0 gm/l and 25 gm/l) are shown in figure 6. Except for the lowest concentration (0.5 gm/l), two distinct fluorescence emission peaks are observed for all other concentrations of RhB in propanol. The wavelength of the first intense peak attributed to the monomer is red-shifted from 603 nm to 635 nm as the concentration of the RhB solution increases. An additional peak around 675 nm is also observed for all the RhB solutions in propanol due to the aggregation of RhB in the solution forming a dimer. With the increase of concentration of the RhB in propanol the intensity of the monomer peak decreases and the dimer peak increases. The intensity of the dimer fluorescence is observed to increase initially with the increase in RhB concentration in the solution, however, with further increases in concentration the peak intensity decreases. In addition to that, for RhB with concentrations 15 gm/l, 18 gm/l and 25 gm/l (inset of figure 6) in the solution, the intensity ratio of monomer to dimer fluorescence peaks is observed to be nearly constant. It can also be seen (inset of figure 6) that there is a small redshift of 15 nm in the dimer fluorescence peak as the concentration of RhB in the solution increases from 12 gm/l to 25 gm/l. The propanol solution with RhB concentration higher than 25 gm/l was found to be saturated at room temperature. It is worthwhile to mention here that no significant changes were observed in the fluorescence spectra of RhB solution of concentration 3.5 gm/l, 4.5 gm/l, 6.5 gm/l, and 10 gm/l. Hence these are not presented in figure 6.

The monomer and dimer fluorescence peak intensity corresponding to different concentrations of RhB along with the monomer to dimer fluorescence peak intensity ratio is shown in table 3. A graph of monomer to dimer fluorescence peak intensity ratio is plotted against the concentration of RhB in propanol is shown in figure 7 and the nature of the graph is found to follow a power equation

$y = 1.96011\ x^{-0.46996}$   with $R^2 = 0.99142$ (3)

where x represents the concentration of RhB in propanol and y represents the monomer to dimer fluorescence emission peak intensity ratio. $R^2$ is a statistical measure which quantifies the goodness of fit of the model.

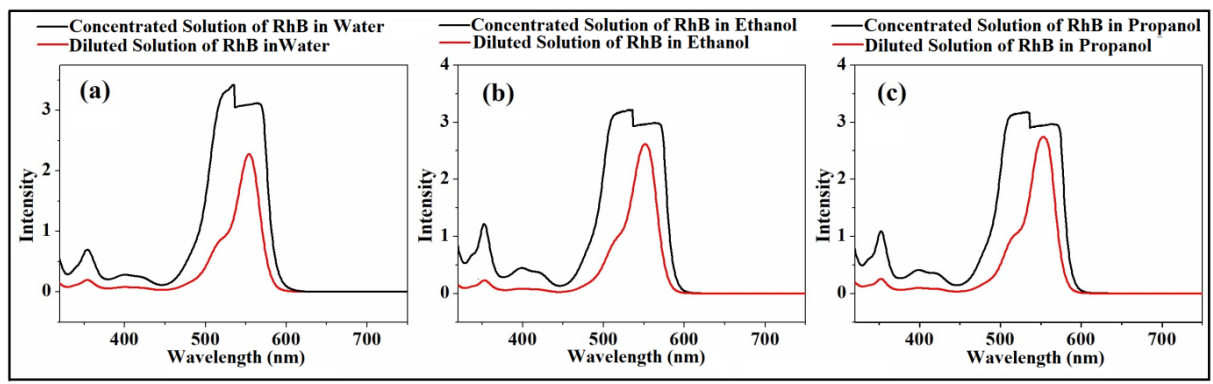

**Figure 1:** Absorption spectra of a low and a high concentrated solution of rhodamine B (RhB) in (a) Water, (b) Ethanol, and (c) Propanol.

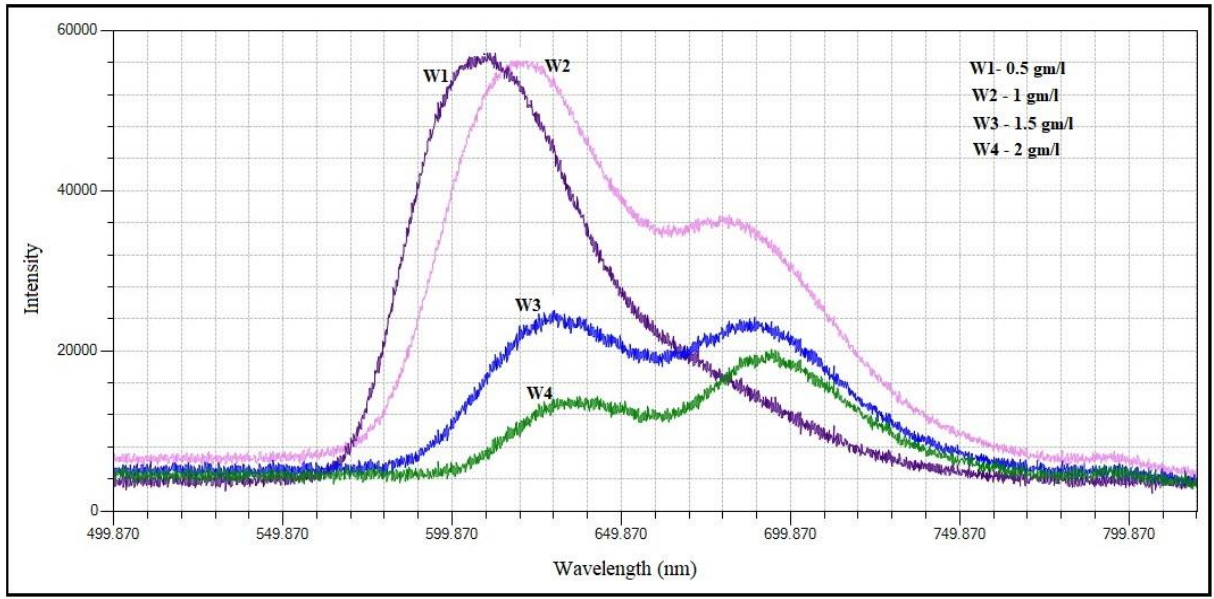

**Figure 2:** Fluorescence emission spectra of the monomeric and dimeric form of rhodamine B (RhB) in water.

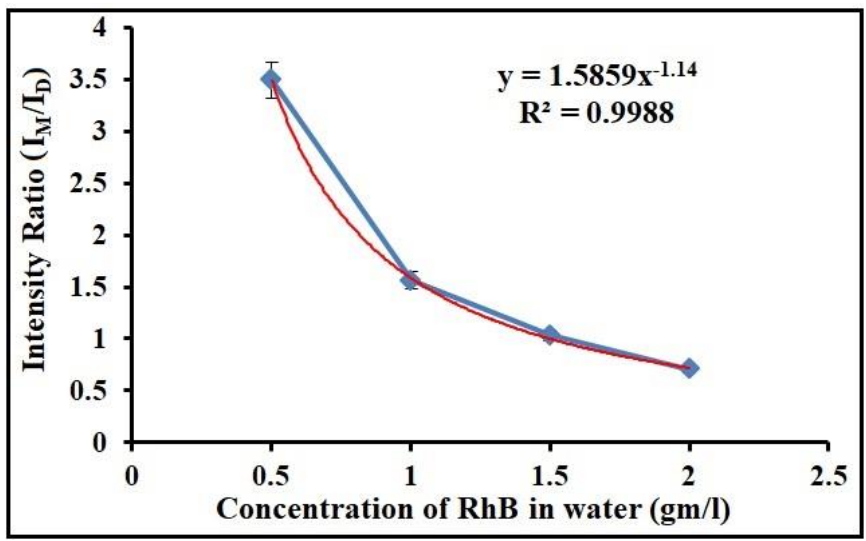

**Figure 3:** Plot of monomer to dimer fluorescence peak intensity ratio $\left(\frac{I_M}{I_D}\right)$ against different concentrations of rhodamine B (RhB) in water. This plot follows a perfect power equation, y = 1.5859 $x^{-1.14}$, with $R^2$ = 0.9988, where x and y represent the concentration of RhB in water and monomer to dimer fluorescence emission peak intensity ratio respectively. The *critical concentration* is found to be equal to 1.5 gm/l (*critical concentration:* The concentration at which the monomer to dimer fluorescence peak intensity ratio $\left(\frac{I_M}{I_D}\right)$ becomes unity, $I_M$= Monomer fluorescence intensity, $I_D$= Dimer fluorescence intensity)

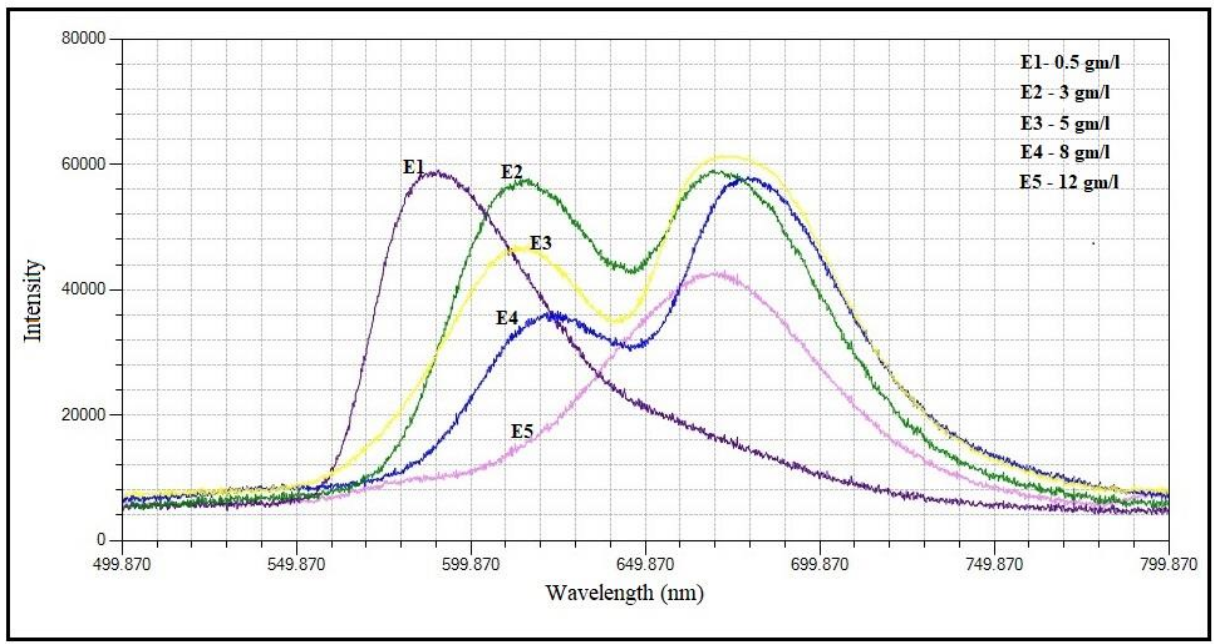

**Figure 4:** Fluorescence emission spectra of the monomeric and dimeric form of rhodamine B (RhB) in ethanol.

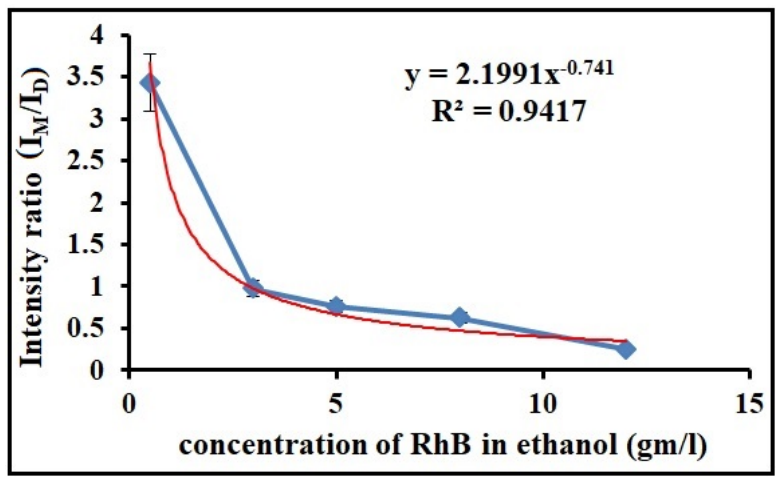

**Figure 5:** Plot of monomer to dimer fluorescence peak intensity ratio $\left(\frac{I_M}{I_D}\right)$ against different concentrations of rhodamine B (RhB) in ethanol. This graph is found to follow a power equation y = 2.1991x$^{-0.741}$, with R² = 0.9417, where x represents the concentration of RhB in propanol and y represents the monomer to dimer fluorescence emission peak intensity ratio. The *critical concentration* is found to be equal to 3 gm/l (*critical concentration:* The concentration at which the monomer to dimer fluorescence peak intensity ratio $\left(\frac{I_M}{I_D}\right)$ becomes unity, $I_M$= Monomer fluorescence intensity, $I_D$= Dimer fluorescence intensity).

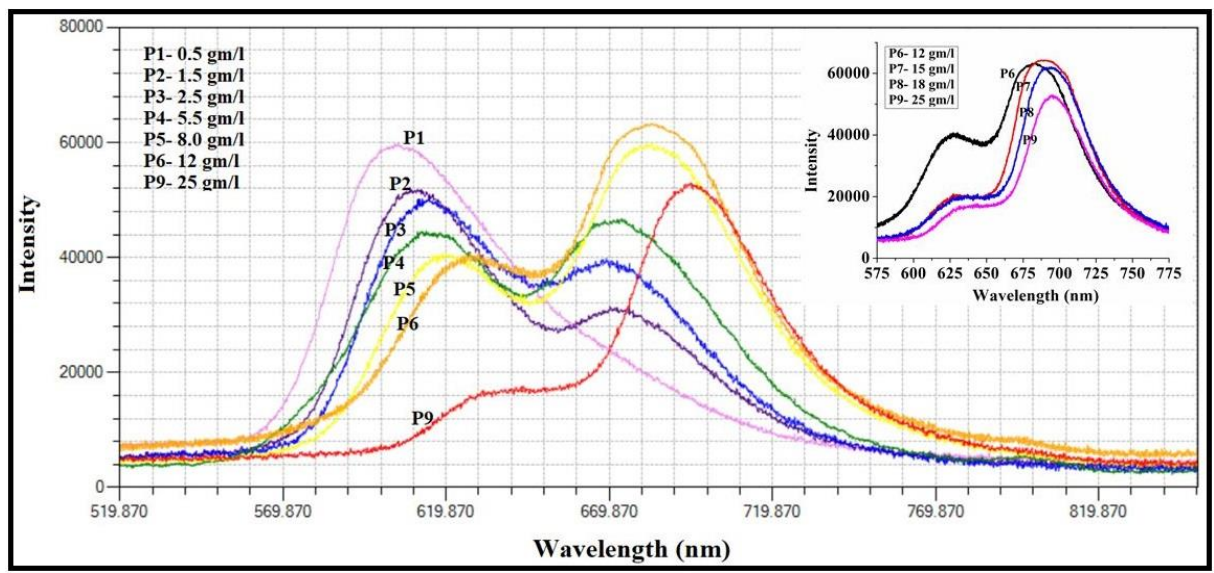

**Figure 6:** Fluorescence emission spectra of the monomeric and dimeric form of rhodamine B (RhB) in propanol.

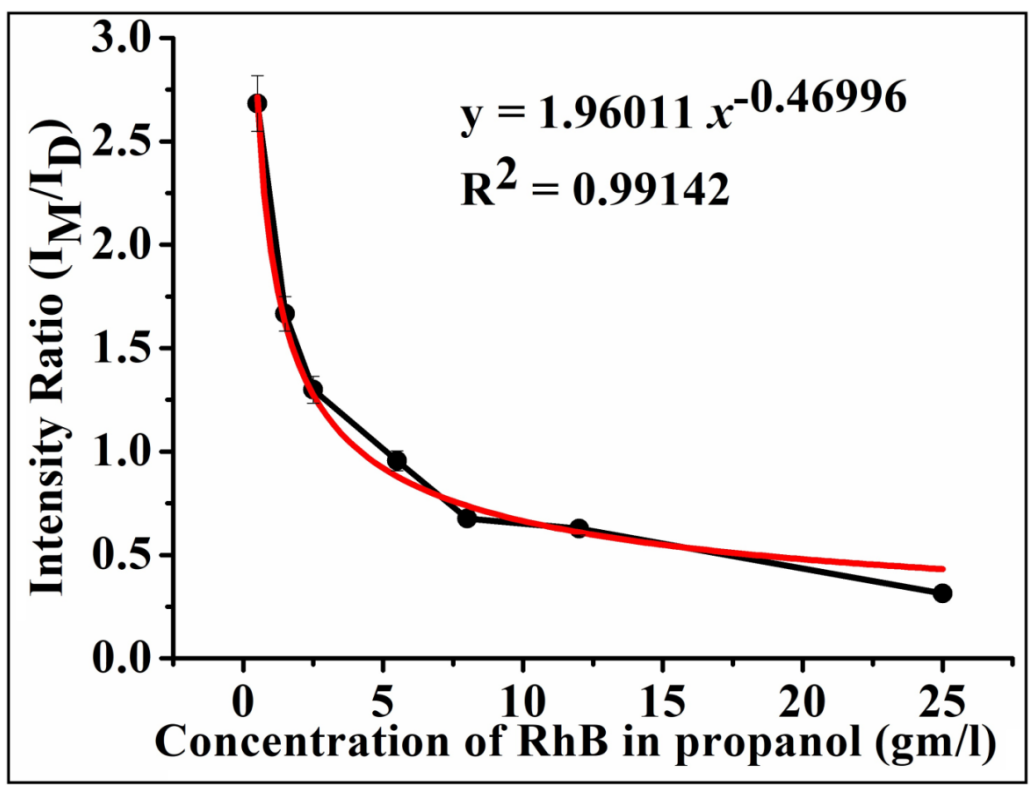

**Figure 7:** Plot of monomer to dimer fluorescence peak intensity ratio $\left(\frac{I_M}{I_D}\right)$ against different concentrations of rhodamine B (RhB) in propanol. This graph is found to follow a power equation $y = 1.96011\, x^{-0.46996}$, with $R^2 = 0.99142$, where x represents the concentration of

RhB in propanol and y represents the monomer to dimer fluorescence emission peak intensity ratio. The *critical concentration* is found to be equal to 4.2 gm/l (*critical concentration:* The concentration at which the monomer to dimer fluorescence peak intensity ratio $\left(\frac{I_M}{I_D}\right)$ becomes unity, $I_M$= Monomer fluorescence intensity, $I_D$= Dimer fluorescence intensity).

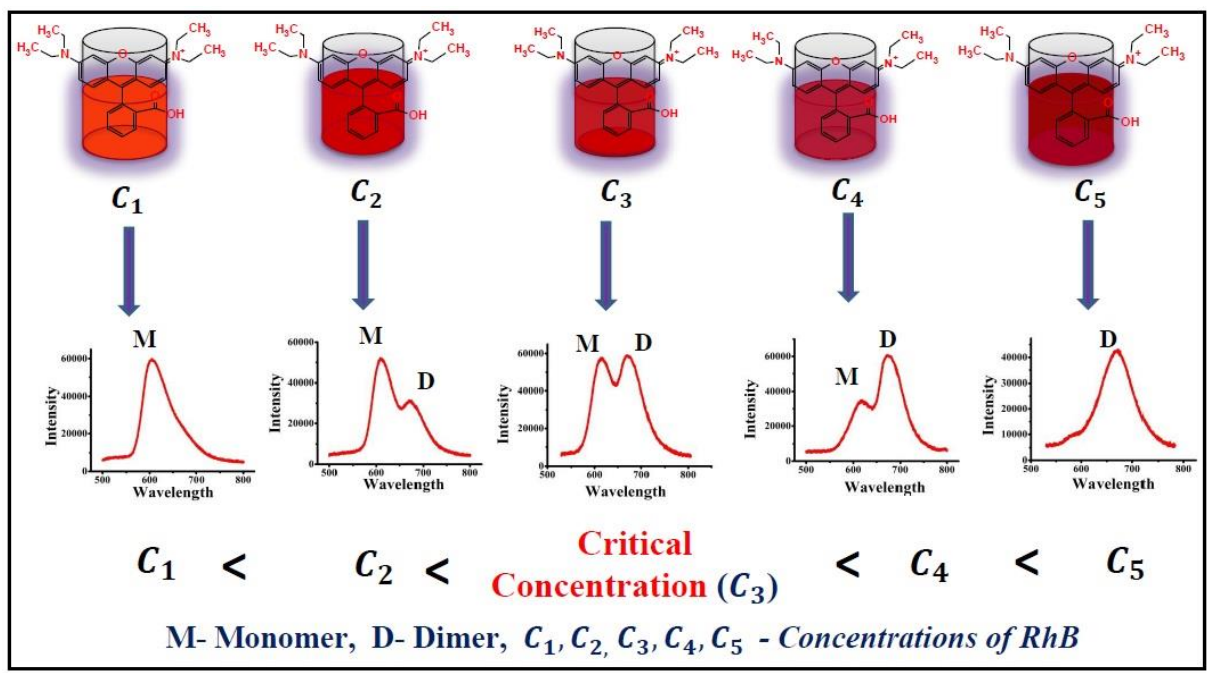

**Figure 8:** Illustrative representation of *critical concentration* of aggregation of dye fluorophore in solvent. At *critical concentration* ($C_3$) the monomer and dimer fluorescence intensity of the fluorophore becomes equal. At concentration ($C_1$, $C_2$) below the *critical concentration,* intensity of monomer fluorescence is more than the dimer fluorescence. At concentration ($C_4$, $C_5$) above the *critical concentration,* the intensity of dimer fluorescence is more than the monomer fluorescence.

**Table 1:** Monomer and dimer fluorescence emission peak intensity of rhodamine-B (RhB) solution in water at different concentration along with their intensity ratio. The intensity of the monomer peak is divided by the intensity of the corresponding dimer peak to evaluate the intensity ratio.

| Concentration of RhB (gm/l: gram/litre) | Monomer intensity ($I_M$) | Dimer intensity ($I_D$) | Ratio $\frac{I_M}{I_D}$ |
|---|---|---|---|
| 0.5 | 57242 | 16384 | 3.494 |
| 1.0 | 55875 | 35724 | 1.564 |
| 1.5 | 24026 | 23256 | 1.033 |
| 2.0 | 12478 | 17662 | 0.706 |

**Table 2:** Monomer and dimer fluorescence emission peak intensity of rhodamine-B (RhB) solution in ethanol at different concentration along with their intensity ratio. The intensity of the monomer peak is divided by the intensity of the corresponding dimer peak to evaluate the intensity ratio.

| Concentration of RhB (gm/l: gram/litre) | Monomer intensity ($I_M$) | Dimer intensity ($I_D$) | Ratio $\frac{I_M}{I_D}$ |
|---|---|---|---|
| 0.5 | 58270 | 16992 | 3.45 |
| 3.0 | 57249 | 58836 | 0.97 |
| 5.0 | 46624 | 61084 | 0.76 |
| 8.0 | 35584 | 57515 | 0.62 |
| 12.0 | 10360 | 42246 | 0.25 |

**Table 3:** Monomer and dimer fluorescence emission peak intensity of Rhodamine-B (RhB) solution in propanol at different concentration along with their intensity ratio. The intensity of the monomer peak is divided by the intensity of the corresponding dimer peak to evaluate the intensity ratio.

| Concentration of RhB (gm/l: gram/litre) | Monomer intensity ($I_M$) | Dimer intensity ($I_D$) | Ratio $\frac{I_M}{I_D}$ |
|---|---|---|---|
| 0.5 | 59769 | 22280 | 2.683 |
| 1.5 | 51241 | 30762 | 1.666 |
| 2.5 | 49900 | 38415 | 1.299 |
| 5.5 | 44128 | 46222 | 0.955 |
| 8.0 | 40000 | 59144 | 0.676 |
| 12.0 | 39648 | 63076 | 0.628 |
| 25.0 | 16350 | 52173 | 0.313 |

*Aggregation behaviour of rhodamine B (RhB)*

The simultaneous observation of monomer and dimer fluorescence is essential to evaluate the trend of their intensity ratio with respect to the concentration of the solute in different solvents. In this experiment, it has been found that the monomer to dimer fluorescence peak intensity ratio for all the solvents (water, ethanol, and propanol) decreases as the concentration of the solute (RhB) increases. The decrease of monomer fluorescence intensity with concentration may be due to long range dipole-dipole energy transfer from monomer excited state to the aggregates [27]. This mechanism may also be affected by excitation energy transfer between monomers [28]. The rate of self-aggregation of RhB in the solution will also increase with the increase of its concentration and that increases the dimer fluorescence peak intensity. The self-aggregation or the formation of a dimer of RhB in all the solvents is confirmed by the absorption spectrum (figure 1). The monomer and dimer fluorescence emission were simultaneously observed for different ranges of concentration of

RhB in the solvents. In water, it is the smallest (0.5 gm/l to 2 gm/l) and in propanol, it is the highest (0.5 gm/l to 25 gm/l). The intensity of the dimer fluorescence peak in water is very less than the corresponding peak in ethanol and propanol. The ethanol solution at the highest concentration of RhB (12 gm/l) possesses the highest quantum yield of dimer fluorescence as the monomer fluorescence is almost missing.

Being a cationic dye [29] the force between two RhB dye molecules is coulombic repulsive in nature. The coulombic force depends on the dielectric constant of the medium where the charge species exist. Coulombic force of attraction or repulsion reduces with the increase in the dielectric constant of the medium. Here the RhB dye is dissolved in different solvents with different dielectric constants. Thus the extent of the repulsive force between two RhB molecules is different in different solvents. The dielectric constant is the highest in water (80.10) followed by ethanol (25.10) and propanol (17.9) [30]. Therefore, the magnitude of the repulsive force is less in water than in ethanol and propanol. As a result, in water RhB molecules can easily overcome the repulsive force and approach one another, which favour the aggregation process [31]. Similarly, the dependency of the aggregation behaviour of the RhB molecules in different solvents with different dielectric constants can be explained. In general, it can be concluded that the solvents with higher dielectric constant help the RhB molecules to form the aggregates by reducing the electrostatic repulsive force between the cationic RhB dye molecules.

The plot of monomer to dimer fluorescence peak intensity ratio against the concentration of RhB follows a specific trend in different solvents (Equation 1, 2 and 3). It is clear from the equations that for a certain concentration of RhB in any solvent (water, ethanol and propanol) the magnitude of the ratio $\left(\frac{I_M}{I_D}\right)$ is different and it signifies the strength of aggregation of the solute (RhB) molecule in the solvent. It is interesting to observe that in water the intensity of the monomer and dimer fluorescence is almost equal for 1.5 gm/l concentration of RhB (figure 3,) and hence their ratio is equivalent to one. In the same way, the ratio of monomer to dimer fluorescence peak intensity becomes equivalent to one for 3 gm/l (figure 5) and 4.2 gm/l (figure 7) concentration of RhB in ethanol and propanol respectively. We propose a term for this concentration value of RhB at which the ratio $\left(\frac{I_M}{I_D}\right)$ become equivalent to one as "*critical concentration*". Above this critical concentration, the dimeric fluorescence dominates and below which the monomeric fluorescence dominates. A generalised concept of the *critical concentration* can be visualised graphically from the figure

8. The parameter "*critical concentration*" will provide a condition of equilibrium between the monomer and their aggregates without knowing the actual ratio of different conformations of the fluorophore leading to the different fluorescence peaks. It is observed that, the solvent for which the magnitude of *critical concentration* is less the fluorophore molecule aggregates more easily and vice versa. Thus, in this experiment, water is the most suitable solvent for RhB molecules for self-aggregation than ethanol and propanol but with less dimeric quantum yield. The *critical concentration* is found to be the lowest in water and the highest in propanol, whereas the dielectric constant is highest in water and lowest in propanol. So, the *critical concentration* varies inversely with the dielectric constant of the solvent. Altogether, it can be said that the ratio $\left(\frac{I_M}{I_D}\right)$ between the monomer and dimer fluorescence peak intensity and *critical concentration* quantify the aggregation process of the fluorophore in the solvent. Hence, the "*critical concentration*" value of the fluorophore (RhB) could be assumed as the characteristic parameter to study the aggregation behaviour of the fluorophore in solvents.

**Conclusion:**

This work provides a unique method of characterization of the aggregation behaviour of RhB in water and alcohol (ethanol and propanol). The intensity ratio of the monomer to dimer fluorescence peak of the fluorophore or the *critical concentration* as the characteristic parameter of the aggregation behaviour is advantageous over the absorption studies. In this process, the aggregation behaviour of the fluorophore can be quantified up to its saturation point in the solvent. The definition of the parameter "*critical concentration*" based on the fluorescence intensity ratio of monomer and their aggregates; hence it is free from the effect of any other experimental parameter which can influence the intensity of any fluorescence peak. The RhB is generally dissolved in the different solvents [13] to use as laser gain media. Therefore the lasing ability and the wavelength tunability of the gain media could be enhanced if the fluorescence emission characteristics of different aggregates of the fluorophore are known. Further, the presence of a foreign molecule (example: heavy metal) in the solvent (example: water) along with the fluorophore could affect its fluorescence characteristics. Under such conditions, there is a possibility of a change of magnitude of the monomer to dimer fluorescence peak intensity ratio. This change in the fluorescence spectra may be used as a specific method of detection of that particular foreign molecule in the solvent. Hence, this method may have the merit of use in the field of heavy metal sensing in water. The described method of characterization of the aggregation behaviour based on monomer to dimer fluorescence peak intensity ratio may also be extended for other fluorophores and solvents.

**Authors Declarations:**

**Funding:**

This work was supported by Science and Engineering Research Board, Department of Science and Technology, Government of India (Project No. CRG/2021/002942).

**Conflicts of interest/Competing interests:**

The authors declare that they have no competing interests with respect to the research, authorship, and/or publication of this article.

**Ethics approval/declarations**

NA

**Consent for publication**

The authors provide their consent to publish this article in the *Laser Physics*

**Availability of data and materials:**

The data generated and analysed during the present study are available from the corresponding author upon reasonable request.

**Author contribution**

**Simanta Hazarika:** Conceptualisation, Methodology, Investigation, Supervision, Formal Analysis, Visualisation, Writing-original Draft. **Seikh Mustafa Radiul:** Investigation, Formal Analysis, Visualisation, Writing-original Draft. **Jugal Chowdhury**: Methodology, Formal Analysis, Writing-original Draft. **Angana Goswami:** Investigation, Formal Analysis.

**Acknowledgement:**

The authors thank Science and Engineering Research Board, Department of Science and Technology, Government of India for the grant, No. CRG/2021/002942.The authors express sincere thanks to Prof. Buddhadeb Bhattacharjee, Department of Physics, Gauhati University. SH wishes to thank FIST (DST) grant SR/FST/PSI-213/2016(C).